# Relative Performance of a Multi-level Cache with Last-Level Cache Replacement: An Analytic Review

Bijay K.Paikaray

Dept. of CSE, CUTM

Bhubaneswer, India

Debabala Swain

Dept. of CSE, CUTM

Bhubaneswer, India

**Abstract**

Current day processors employ multi-level cache hierarchy with one or two levels of private caches and a shared last-level cache (LLC). An efficient cache replacement policy at LLC is essential for reducing the off-chip memory transfer as well as conflict for memory bandwidth. Cache replacement techniques for inclusive LLCs may not be efficient for multilevel cache as it can be shared by enormous applications with varying access behavior, running simultaneously. One application may dominate another by flooding of cache requests and evicting the useful data of the other application. From the performance point of view, an exclusive LLC make the replacement policies more demanding, as compared to an inclusive LLC. This paper analyzes some of the existing replacement techniques on the LLC with their performance assessment.

**Keywords:** Cache Performance, Multi-Level Cache, Last Level Cache (LLC).

## 1. Introduction

A cache memory is mainly designed to alleviate the speed gap between the processor and main memory, which is determined by the access time and miss rates. Caching was initially performed by a single cache on a uniprocessor, but as the processor applications and their competition increased, multi-level cache and multi-core processors were necessary in computer systems. The first level (L1 cache) is closest to the processor and it is used for direct access, while the second or the third level (L2 cache, L3 cache) is referred to as LLC (Last-Level Cache) are designed to hide long miss penalty of accessing to main memory, which involves hundreds of processor cycles in current microprocessors.

The system performance strongly depends on the cache hierarchy performance. Thus, many researches have been done to improve the cache performance, although usually focusing on a given level of the cache hierarchy (i.e., L1, L2 or L3). Techniques like load-bypassing, way-prediction, or prefetching, have been widely investigated and implemented in many commercial products. Although these techniques have been successfully implemented in typical monolithic processors, but the performance of the LLC in particular, is a major design issue in current processors.

Now a day's large size LLC are designed to keep much information and reduce capacity misses. So the sizes range from several hundreds of KB up to several MB [3] [4]. In order to keep low number of conflict misses, current LLCs implement a large number of ways (like 16 ways/32 ways). Mostly, caches use temporal locality by implementing the Least Recently Used (LRU) replacement algorithm.

This algorithm acts as a stack that places the most recent block on the top of the stack and the least recent block on the bottom, which is the block to be evicted during replacement.

**2. Literature Review**

Different cache eviction policies have been proposed in this section for both unicore and multicore processors. The system performance strongly depends on the cache hierarchy performance. Although it usually focus on a given level of the cache hierarchy (i.e., L1, L2 or L3). Techniques like load-bypassing, way-prediction, or prefetching, have been widely investigated and implemented in many commercial products. Though these techniques have been successfully implemented in typical monolithic processors, the pressure on the memory controller is much higher in multicore and manycore systems than in monolithic processors. Therefore, the performance of the cache hierarchy in general, and the performance of the LLC in particular, is a major design concern in current processors.

Caches in the first-level (i.e. in L1) use to be very small as compared to other levels, so it is important to efficiently handle the available space. Some schemes utilize information of past behavior of a given block, using reuse information, to improve the cache performance. The NTS approach by Tyson et al. [5] and the MAT proposed by Johnson et al. [6] are some of the examples. The former marks a block as cacheable based on its reuse information, the latter classifies the blocks as temporal and not temporal based on their reuse information during its past live time. Rivers et al. in [7] propose to exploit reuse information based on the effective address of the referenced data and on the program counter of the load instruction.

In [8], Lin and Reinhardt proposed a hardware approach that predicts when to evict a block before it reaches the bottom of the LRU stack. The first approach is referred to as *sequence-based* prediction. This approach records and predicts the sequence of memory events leading up to the *last touch* of a block. The second one is the *time-based* approach, which tracks a line's timing to predict when a line's last touch will likely occur. In [9], Kharbutli et al. proposed the counter-based L2 cache replacement, which predicts when to evict a block before it reaches the bottom of the stack. Two approaches are presented for the way prediction i.e. Access Interval Predictor (AIP) and the Live-time Predictor (LvP) [10].

The AIP scheme predicts by using counters to keep track of the number of accesses to the same set during a given access interval of a cache line. If the counter reaches a threshold value, the associated line can be selected for replacement. In LvP it counts the number of accesses to each line instead of to the same set.

**3. Recently Proposed Schemes**

This section represents the research works related with the placement and replacement schemes at the Last Level Cache (LLC), along with their performance analysis.

*3.1. Replacement Schemes in Unicore Processors*

Counter-based replacement technique [9] for unicore LLC predicts access interval using a counter for each cache line. All counters in a set are incremented on an access and a cache line whose counter value exceeds a given threshold is selected as victim.

The use of reuse information during victim selection for unicore LLC has been exploited in [11]. It predicts the reuse distance and uses the predicted values for cache eviction. On a cache miss, if the predicted reuse distance of the memory reference is higher than the reuse distance seen by all cache lines in the set, the requested data is directly sent to the processor without storing it in the cache. Otherwise, a cache line with highest reuse distance is replaced with the requested data. Using Adaptive Weight Ranking Policy [12], ranking of the referenced blocks is done three factors. $F_i$, $R_i$, $N$, where $F_i$ be the frequency index which holds the frequency of each block $i$ in the buffer and $R_i$ be the recency index which shows the clock of last access when buffer has been referenced and $N$ is the total number of access to be made. Then the weighting value of block $i$ can be computed as:

$$W_i = \frac{F_i}{N - R_i} \quad \text{---[12]}$$

*3.2 Replacement Schemes in Multicore Processors*

**3.2.1 MRUT-based Prediction and Replacement [10]**

The MRU-based algorithms based on the concepts of live and dead times of a block have been widely used in cache research [13]. The generation time of a block defines the elapsed time since the block is fetched into the cache until it is replaced. This amount of time can be divided in live and dead times. The live time refers to the elapsed time since the block is fetched until its last access before it is replaced, and the dead time refers to the time from its last access until eviction. Figure 1 illustrates the concept of MRUT in the context of the live time of the *A* cache block, assuming the LRU replacement policy.

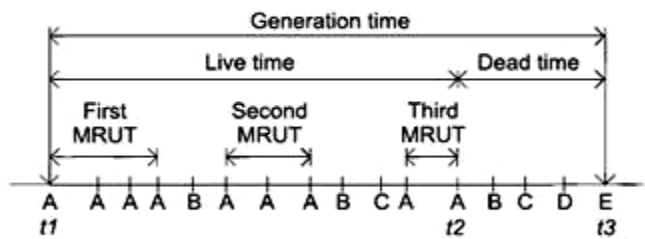

**Figure 1. Generation time of the A cache block** [10].

Initially block *A* is allocated at time $t_1$ in the MRU position at the top of the LRU stack. The block maintains this position while it is being accessed. Then, the block leaves this position because block *B* is accessed. At this point, we say that block *A* has finished its first MRUT.

After accessing block *B*, block *A* is referenced again so returning to the MRU position and starting a second MRUT. At time $t_2$, block *A* finishes its third MRUT, which is the last MRU-Tour of this block before leaving the cache at time $t_3$. MRUT-based replacement algorithms can be a single or multiple MRUTs depending on the number of times the blocks are evicted. Figure 2 depicts the results for the SPEC2000 benchmark suite [14] and 1MB-16way L2 cache under the LRU algorithm.

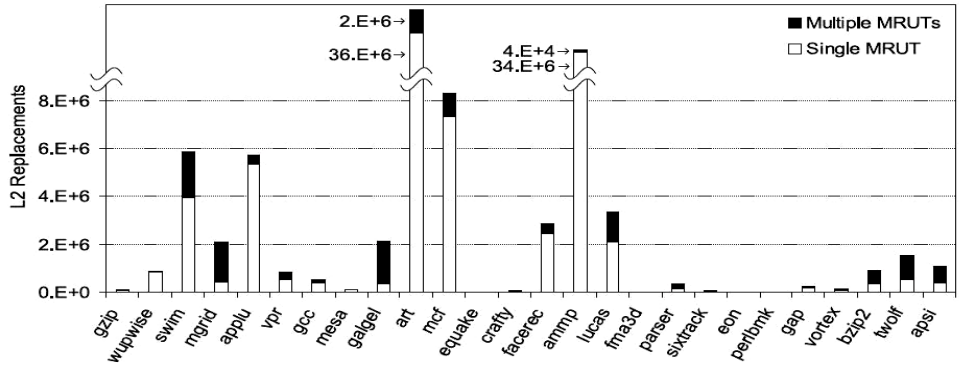

**Figure 2. Number of replacements split into single and multiple MRUTs[10].**

### 3.2.2 The Application-Aware Cache Replacement [1]

This policy prevents victimizing low-access rate application by a high-access rate application. To make this policy access rate aware, separate local eviction priority chains are maintained for different cores. ACR technique updates the eviction priority of only those cache lines that belong to the referencing application and maintains separate eviction priority chains for each application.

It dynamically keeps track of maximum life-time of cache lines in shared LLC for each concurrent application and helps in efficient utilization of the cache space. Experimental evaluation of ACR technique for 2-core and 4-core systems using SPEC CPU 2000 and 2006 benchmark suites shows significant speed-up improvement over the *least recently used* and *thread-aware dynamic re-reference interval prediction* techniques. Each application has different reuse patterns, which can change during various stages of the execution.

| SPEC benchmark | LLC | statistics | SPEC benchmark | LLC | statistics |
|---|---|---|---|---|---|
| | APKI | Miss Rate (%) | | APKI | Miss Rate (%) |
| 164.gzip | 1.22 | 17.08 | 429.mcf | 64.47 | 90.91 |
| 168.wupwise | 3.01 | 99.13 | 435.gromacs | 1.72 | 19.59 |
| 171.swim | 22.89 | 99.98 | 437.leslie3d | 9.15 | 82.64 |
| 172.mgrid | 12.32 | 64.95 | 444.namd | 0.68 | 98.68 |
| 173.applu | 20.16 | 99.92 | 450.soplex | 2.94 | 35.67 |
| 175.vpr | 11.78 | 27.48 | 454.calculix | 0.91 | 62.92 |
| 177.mesa | 0.72 | 91.53 | 456.hmmer | 2.14 | 71.36 |
| 178.galgel | 14.09 | 43.91 | 458.sjeng | 0.37 | 79.98 |
| 179.art | 129.64 | 78.81 | 459.GemsFDTD | 0.006 | 70.98 |
| 186.crafty | 0.58 | 9.65 | 462.libquantum | 6.72 | 99.64 |
| 193.fma3d | 0.00051 | 100 | 464.h264ref | 0.88 | 10.41 |
| 300.twolf | 15.24 | 32.37 | 470.lbm | 32.07 | 99.99 |
| 401.bzip2 | 5.18 | 43.57 | | | |

**Table 1. LLC statistics for SPEC CPU 2000 and 2006 bencmarks[14]**

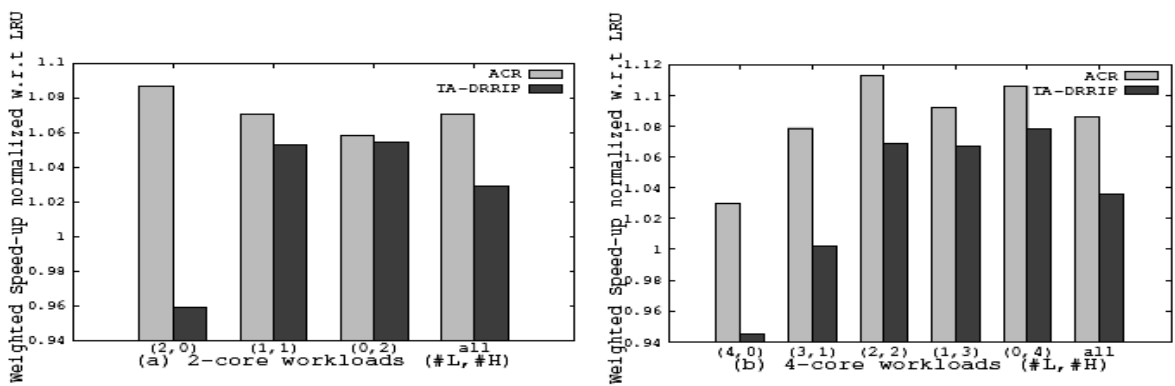

**Figure 3. Comparative performance analysis between ACR & IA-DRRIP for different Workloads[3].**

**Conclusion**

In this paper we presented different cache management mechanisms that determine the cache placement & replacement policies on a per-block basis. The key idea is to predict a missed block's temporal locality before inserting it into the cache and choose the appropriate insertion policy for the block. In our future work we will include developing and analyzing other temporal locality prediction schemes and also investigating the interaction of our mechanism with pre-fetching. We will practical, a low-overhead implementation based scheme on real time simulation using different benchmarks for significant performance compared to other similar approaches on a wide variety of workloads and system configurations. It will be tested for both unicore and multicore processors.